\documentclass[
reprint,
aps,
pra,
superscriptaddress,
longbibliography,
]{revtex4-2}

\usepackage{amsmath}
\usepackage{graphicx}
\usepackage{hyperref}
\hypersetup{
  colorlinks=true,
  urlcolor=blue,
  linkcolor=blue,
  citecolor=blue
}

\usepackage{braket}
\usepackage{xcolor}
\newcommand{\cnuc}[0]{$^{13}$C }
\newcommand{\sinuc}[0]{$^{29}$Si }
\newcommand{\affiliationharvardengineering}[0]{
  \affiliation{
    John A.\ Paulson School of Engineering and Applied Sciences,
    Harvard University,
    Cambridge, MA 02138, USA
  }
}
\newcommand{\affiliationharvardphysics}[0]{
  \affiliation{
    Department of Physics,
    Harvard University,
    Cambridge, MA 02138, USA
  }
}
\newcommand{\affiliationdelft}[0]{
  \affiliation{
    QuTech,
    Delft University of Technology,
    2600 GA Delft, The Netherlands
  }
  \affiliation{
    Kavli Institute of Nanoscience Delft,
    Delft University of Technology,
    2600 GA Delft, The Netherlands
  }
}

\begin{document}

\title{Coherent coupling of mechanics to a single nuclear spin}

\author{Smarak Maity}
\thanks{These authors contributed equally.}
\affiliationharvardengineering

\author{Benjamin Pingault}
\thanks{These authors contributed equally.}
\affiliationharvardengineering
\affiliationdelft

\author{Graham Joe}
\affiliationharvardengineering

\author{Michelle Chalupnik}
\affiliationharvardphysics

\author{Daniel Assump\c{c}\~{a}o}
\author{Eliza Cornell}
\author{Linbo Shao}

\author{Marko Lon\v{c}ar}
\thanks{loncar@seas.harvard.edu}
\affiliationharvardengineering

\begin{abstract}
  Nuclear spins interact weakly with their environment.
  In particular, they are generally insensitive to mechanical vibrations.
  Here, we successfully demonstrate the coherent coupling of mechanics to a single nuclear spin.
  This coupling is mediated by a silicon vacancy (SiV) centre in diamond, taking advantage of its large strain susceptibility and hyperfine interaction with nuclear spins.
  Importantly, we demonstrate that the nuclear spin retains its excellent coherence properties even in the presence of this coupling.
  This provides a way to leverage nuclear spins as quantum memories for mechanical systems in the quantum regime.
\end{abstract}

\maketitle

Mechanical systems are at the forefront of investigations in quantum physics \cite{Safavi-Naeini2013Nature, Aspelmeyer2014RMP, MacCabe2020Science, Delord2020Nature}.
Examples include the realisation of quantum states of mechanical resonators \cite{OConnell2010Nature}, quantum squeezing of mechanical motion \cite{Wollman2015Science}, and coupling of mechanics to other quantum systems \cite{Arcizet2011NatPhys, Kolkowitz2012Science, Palomaki2013Science, Barfuss2015NatPhys, Chu2017Science, Lee2017JOptics, Whiteley2019NatPhys, Arrangoiz-Arriola2019Nature, Delord2020Nature, Maity2020NatComm, Mirhosseini2020Nature}.
However, their size induces significant coupling to their environment, which limits their coherence \cite{Meenehan2015PRX, Satzinger2018Nature, MacCabe2020Science}, and reaching coherence times $\sim 100\text{ $\mu$s}$ requires complex phononic shielding \cite{MacCabe2020Science}.
On the other hand, nuclear spins display coherence times several orders of magnitude longer, the longest among all solid-state systems \cite{Bradley2019PRX, Bartling2021Arxiv}, making them desirable quantum memories \cite{Kalb2017Science, Bradley2019PRX, Pompili2021Science}.
This is due to their insensitivity to the environment, in particular to lattice vibrations.
In this work, we achieve coherent coupling of mechanical vibrations to an individual \cnuc nuclear spin in diamond.
We utilize the large strain susceptibility \cite{Meesala2018PRB, Sohn2018NatComm} and the hyperfine coupling \cite{Pingault2017NatComm, Metsch2019PRL, Nguyen2019PRB} of the negatively charged silicon-vacancy (SiV) centre to mediate this interaction.
We confirm that the nuclear spin retains its excellent coherence properties, thus demonstrating the coherent coupling of mechanics to long-lived quantum memories.

\begin{figure}
  \includegraphics{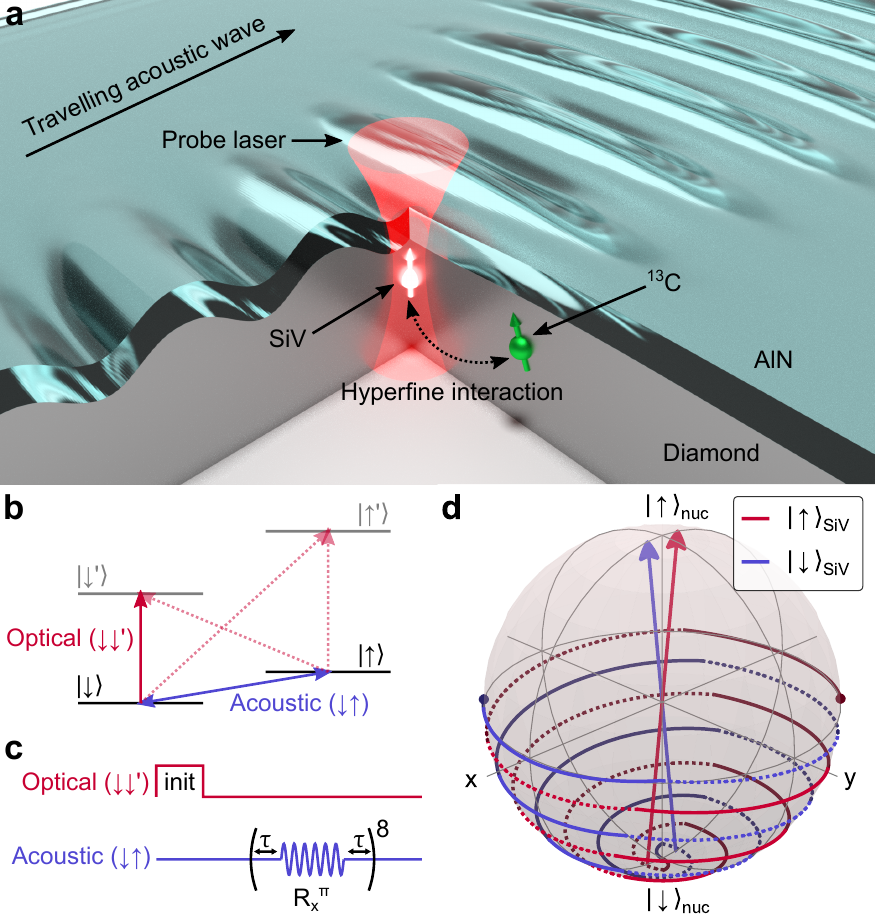}
  \caption{
    \label{fig1}
    \textbf{Principle behind mechanical coupling to nuclear spins.}
    \textbf{a} Schematic of the system.
    Surface acoustic waves (SAWs) generated in the AlN layer, and extending into diamond, are used to control the \cnuc nuclear spin via its hyperfine coupling to the silicon vacancy (SiV) electron spin, which can be initialized and read out optically.
    \textbf{b} Simplified electronic structure of the SiV in the presence of an external magnetic field.
    The red arrows (both dashed and solid) indicate optical transitions around $737\text{ nm}$.
    The solid red arrow corresponds to the transition $(\downarrow\downarrow')$ used for optical pumping to initialize the SiV state in $\ket\uparrow$.
    The blue arrow corresponds to the $3.42\text{ GHz}$ spin-flipping transition $\downarrow\uparrow$ that can be driven acoustically.
    \textbf{c} Schematic of a sequence that induces conditional precession of the nuclear spin.
    The optical part corresponds to the optical initialization of the SiV spin into $\ket\uparrow$.
    The SiV spin can be initialized in $\ket\downarrow$ by exciting the $\uparrow\uparrow'$ transition.
    The acoustic part corresponds to a series of 8 periodically timed acoustic $\pi$-pulses around the $X$-axis of the Bloch sphere ($\mathcal R_\textbf{x}^\pi$).
    \textbf{d} Simulation of the evolution of the nuclear spin on the Bloch sphere for the sequence described in \textbf{c}.
    The red (blue) arrow corresponds to the nuclear spin precession axis when the SiV is in $\ket\uparrow$ ($\ket\downarrow$).
    The nuclear spin starts in $\ket\downarrow_\text{nuc}$ and follows the red (blue) trajectory when the SiV is initially in $\ket\uparrow$ ($\ket\downarrow$).
    Spin flips of the SiV are indicated by transitions between solid and dashed trajectories.
    The red (blue) dot indicates the final state of the nuclear spin for the SiV initially in state $\ket\uparrow$ ($\ket\downarrow$).
    The hyperfine coupling parameters used are $\{A_\parallel=0.11\text{ MHz}, A_\perp=0.33\text{ MHz}\}$ and the inter-pulse time is $\tau = 0.169\text{ $\mu$s}$.
  }
\end{figure}

In our experiments, the mechanical vibrations are surface acoustic waves (SAWs), which have natural two-dimensional confinement, can be guided on chip, and are easily generated by electrical excitation of interdigital transducers (IDTs) \cite{Maity2020NatComm}.
We pattern aluminium IDTs on top of a layer of piezoelectric aluminium nitride deposited on the surface of a single-crystal diamond.
The diamond sample is ultra pure (nitrogen concentration $<5\text{ ppb}$) and has a natural abundance of the \cnuc isotope ($1.1\%$).
The diamond is implanted with $^{28}\text{Si}^+$ ions and annealed to form SiV centres, which can be optically addressed individually.
The SiVs are located within the strain field of the SAW, which extends about a wavelength ($\sim 3\text{ $\mu$m}$) below the surface of the diamond (Fig.\ \ref{fig1}a).
The sample is cooled to below $200\text{ mK}$ in a dilution refrigerator to increase the lifetime of the SiV electron spin ($S=1/2$) \cite{Sukachev2017PRL}.
An external magnetic field of $0.13\text{ T}$ at an angle of $90^\circ$ with respect to the SiV axis lifts the spin degeneracy and introduces a splitting of $f_{\downarrow\uparrow}=3.42\text{ GHz}$ between the $\ket\downarrow$ and $\ket\uparrow$ spin states (Fig.\ \ref{fig1}b).
The optical transition labelled $\downarrow\downarrow'$ can be excited resonantly by laser pulses to initialize the SiV in $\ket\uparrow$ via optical pumping, while the $\downarrow\uparrow$ transition can be driven acoustically to coherently transfer population between $\ket\downarrow$ and $\ket\uparrow$ \cite{Maity2020NatComm}.

The mechanical coupling to the \cnuc nuclear spin ($S=1/2$) relies on the simultaneous coupling of an SiV electronic spin to acoustic waves and to the nuclear spin.
An individual \cnuc nuclear spin couples to an SiV spin via the hyperfine interaction Hamiltonian
\begin{equation*}
  \mathcal H
  = 2\pi\hbar S_z^\text{SiV} \left(
    A_\parallel S_z^\text{nuc}
    + A_\perp S_x^\text{nuc}
  \right),
\end{equation*}
where $S_i = \sigma_i/2$ are the spin operators, with $\sigma_i$ being the Pauli matrices for $i\in\{x,y,z\}$ (``SiV'' and ``nuc'' indicate SiV and nuclear spins respectively), and $A_\parallel$ and $A_\perp$ are the parallel and perpendicular components of the hyperfine coupling.
This coupling causes the nuclear spin to precess about slightly different axes depending on the state of the SiV spin.
This can be utilized to control the nuclear spin state \cite{Taminiau2012PRL, Taminiau2014NatNano, Nguyen2019PRB} with the basic sequence depicted in Fig.\ \ref{fig1}c.
After the SiV spin is initialized in $\ket\uparrow$, a series of periodically timed acoustic pulses (with inter-pulse delay $\tau$) flip its state repeatedly, causing the nuclear spin to alternate between two different precession axes.
If the acoustic pulses are synchronized with the precession of the nuclear spin, such that \cite{Taminiau2014NatNano, Nguyen2019PRL}
\begin{equation*}
  \tau \approx \frac{2k+1}{4 f_\text{L}}\left[1 - \frac{A_\perp^2}{8 f_\text{L}^2}\right],
\end{equation*}
for integer values of $k$, where $f_\text{L}$ is the nuclear Larmor frequency, the nuclear spin rotates on the Bloch sphere by an angle $\phi$, around an effective axis $\textbf n$.
The angle $\phi$ is determined by $\tau$ and the number of pulses $N$, while the effective rotation axis $\textbf n_\downarrow$ or $\textbf n_\uparrow$ depends on $\tau$ and the initial state ($\ket\downarrow$ or $\ket\uparrow$) of the SiV spin.
Fig.\ \ref{fig1}d depicts the effect of the sequence for the initial states $\ket{\downarrow}_\text{SiV}\ket{\downarrow}_\text{nuc}$ and $\ket{\uparrow}_\text{SiV}\ket{\downarrow}_\text{nuc}$, where the final state of the nuclear spin depends on the initial state of the SiV spin.
This is a conditional rotation gate, denoted by $\mathcal R^\phi_{\mathbf n_\downarrow, \mathbf n_\uparrow}$, that entangles the SiV and nuclear spins, and is the building block of all the nuclear spin control sequences performed in this work.

\begin{figure}
  \includegraphics{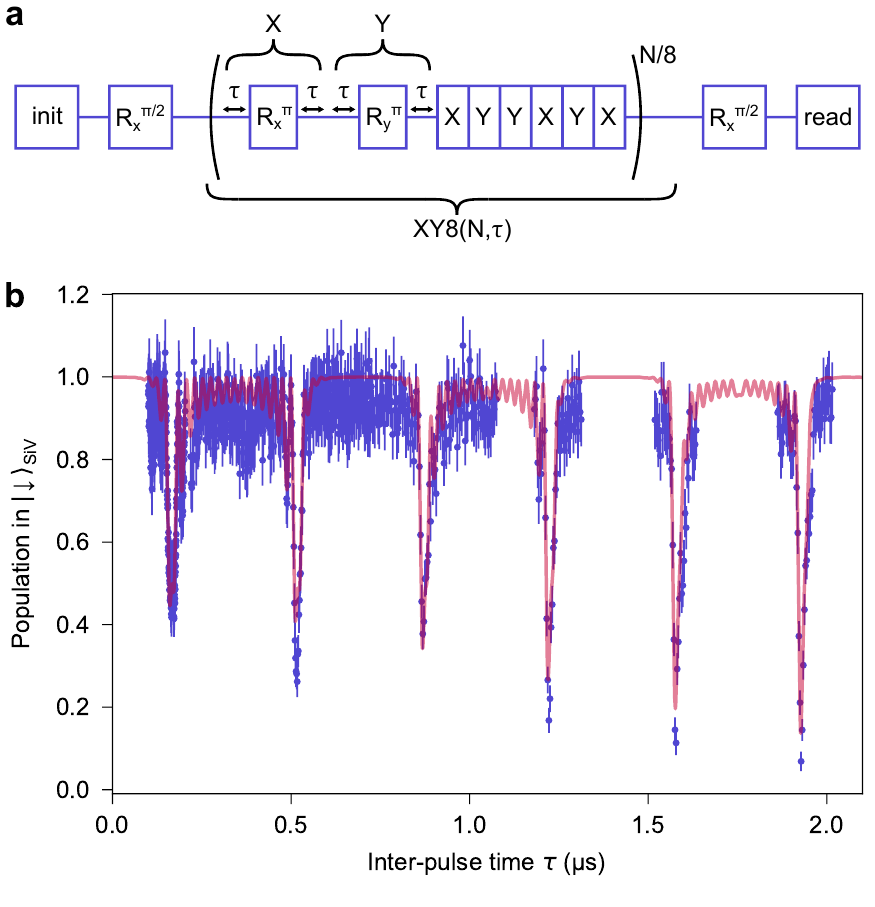}
  \caption{
    \label{fig2}
    \textbf{Nuclear spin resonance spectroscopy using an XY8 decoupling sequence on the SiV spin.}
    \textbf{a} Schematic of the XY8 sequence.
    $\mathcal R_\textbf{x}^\theta$ ($\mathcal R_\textbf{y}^\theta$) represents a pulse that rotates the SiV spin by an angle $\theta$ around the $X$ ($Y$)-axis of the Bloch sphere.
    ``init'' and ``read'' represent optical initialization and readout respectively.
    $N$ is the total number of $\pi$-pulses.
    \textbf{b} Measurement of the final SiV spin population in $\ket\downarrow$ for an XY8 sequence with $N=16$ pulses, as a function of the inter-pulse time $\tau$.
    The blue dots correspond to the experimental data.
    The error bars represent the standard deviation of the measured SiV spin population, from photon counting statistics \cite{Supp}.
    A fit (red curve) to the data yields the hyperfine coupling parameters $\{A_\parallel=0.11\text{ MHz}, A_\perp=0.33\text{ MHz}\}$ for the target nuclear spin \cite{Supp}.
  }
\end{figure}

We identify a target \cnuc spin by performing nuclear spin resonance spectroscopy using the SiV \cite{Taminiau2012PRL, Taminiau2014NatNano}.
The SiV spin is initialized in $\ket\uparrow$, after which an acoustic pulse performs a $\pi/2$ rotation of the spin to bring it into the superposition state $(\ket\uparrow + \ket\downarrow)/\sqrt 2$.
A series of acoustic $\pi$-pulses then perform an XY8 sequence \cite{Ahmed2013PRL}, whereby the SiV spin is flipped repeatedly around two different axes ($X, Y$) on the Bloch sphere, depending on the phase ($0$ or $\pi/2$) of each pulse (Fig.\ \ref{fig2}a).
A final acoustic $\pi/2$ pulse is then applied to the SiV spin, before its state is read out by optical excitation of the $\downarrow\downarrow'$ transition and collection of the resulting fluorescence.
This pulse sequence increases the coherence time of the SiV spin by decoupling it from magnetic field noise \cite{Supp}, while allowing its selective coupling to individual \cnuc nuclear spins.
Figure \ref{fig2}b shows the measurement of the SiV population in $\ket\downarrow$ after an XY8 sequence, as the inter-pulse time $\tau$ is varied.
For values of $\tau$ corresponding to resonance with a particular \cnuc spin, the SiV spin state coherently evolves in conjunction with the nuclear spin.
This appears as periodic sharp dips.
The most prominent dips in Fig.\ \ref{fig2}b can be ascribed to a single nuclear spin, and their positions and depths can be modelled to extract the hyperfine coupling parameters $\{A_\parallel = 0.11\text{ MHz}, A_\perp = 0.33\text{ MHz}\}$ \cite{Supp}.

\begin{figure*}
  \includegraphics{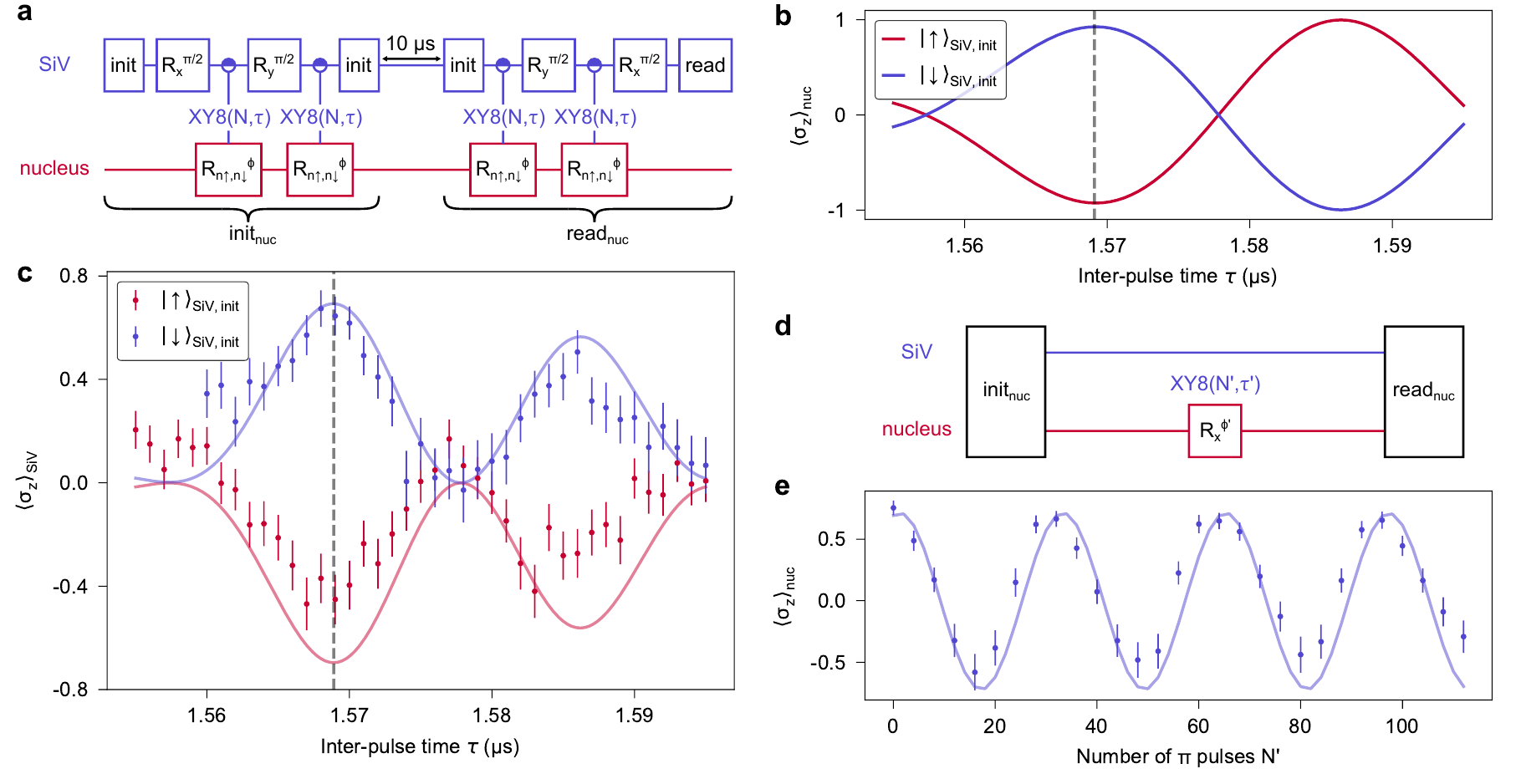}
  \caption{
    \label{fig3}
    \textbf{Mechanical control of a single \cnuc nuclear spin.}
    \textbf{a} Sequences for mechanically driven nuclear spin initialization ($\text{init}_\text{nuc}$) and readout ($\text{read}_\text{nuc}$).
    These are based on entangling conditional rotations of the form $\mathcal R^\phi_{\mathbf{n}\downarrow,\mathbf{n}\uparrow}$ implemented as acoustic XY8 sequences on the SiV spin.
    \textbf{b} Simulation of the state of the nuclear spin after the nuclear initialization sequence for $N=16$ pulses, as a function of the inter-pulse time $\tau$.
    The red (blue) curve corresponds to a sequence at the beginning of which the SiV spin is in the state $\ket\uparrow$ ($\ket\downarrow$).
    The dashed grey line indicates the value of $\tau=1.569\text{ $\mu$s}$ later used for nuclear spin initialization.
    \textbf{c} Measurement (dots) and simulation (solid curves) of the final SiV spin projection after the nuclear initialization and readout sequences, as a function of $\tau$, for different initial states of the SiV spin.
    The measured optimal value of $\tau \approx 1.569\text{ $\mu$s}$ (dashed grey line) agrees with simulations.
    The simulation includes the effect of a second non-resonant nuclear spin with weaker hyperfine couplings, which accounts for the reduction in fidelity \cite{Supp}.
    \textbf{d} Sequence for coherent control of the nuclear spin.
    After nuclear initialization with $\tau=1.569\text{ $\mu$s}$, an XY8 sequence with $\tau'=1.578\text{ $\mu$s}$ rotates the nuclear spin around the $X$-axis of the Bloch sphere with an angle $\phi'$ depending on the number of $\pi$ pulses ($N'$).
    \textbf{e} Mechanically driven Rabi oscillations of the nuclear spin, corresponding to the sequence in \textbf{d}.
    The dots represent measured values, while the solid curve corresponds to a simulation using previously determined hyperfine coupling parameters.
    The error bars in \textbf{c} and \textbf{e} represent the standard deviations of the measured quantities.
  }
\end{figure*}

Having identified the target nuclear spin, we proceed to initialize it.
The sequence for nuclear spin initialization is shown in Fig.\ \ref{fig3}a.
It consists of $\pi/2$ rotations of the SiV spin and conditional rotations of the nuclear spin implemented as XY8 sequences on the SiV spin \cite{Taminiau2012PRL, Taminiau2014NatNano, Nguyen2019PRB}.
We simulate the state of the nuclear spin at the end of the initialization sequence, for values of the inter-pulse delay $\tau$ in the vicinity of the $\tau = 1.578\text{ $\mu$s}$ resonance, as shown in Fig.\ \ref{fig3}b.
For $\tau=1.569\text{ $\mu$s}$, if the SiV is set to $\ket\uparrow_\text{SiV}$ ($\ket\downarrow_\text{SiV}$) at the beginning of the sequence, the nuclear spin is initialized in $\ket\downarrow_\text{nuc}$ ($\ket\uparrow_\text{nuc}$) at the end of it.
The nuclear spin readout sequence corresponds to the same sequence in reverse, and transfers the nuclear spin onto the SiV spin.
The experimental results for the full sequence for various values of $\tau$ are shown as dots in Fig.\ \ref{fig3}c, with simulations shown as solid curves.
For $\tau=1.569\text{ $\mu$s}$, the final state of the SiV depends on the state the nuclear spin was initialized in.
The experimental data agree with the expected results from the simulation, thus confirming the successful initialization and readout of the nuclear spin.
The parasitic partial addressing of other \cnuc nuclear spins with similar hyperfine couplings to the SiV limits the fidelity of the initialization and readout sequences to $0.91$ \cite{Supp}.
The experimentally measured fidelities are $0.85\pm 0.03$ and $0.91\pm 0.02$ for initialization into the $\ket\downarrow_\text{nuc}$ and $\ket\uparrow_\text{nuc}$ states respectively, with readout fidelities equal to the initialization fidelities.
Although a similar initialization sequence could be constructed with $\tau = 1.586\text{ $\mu$s}$, its fidelity is inferior as seen in Fig.\ \ref{fig3}c, due to parasitic partial addressing of other nuclear spins.
The fidelity can be improved by operating at longer inter-pulse delay times, to better resolve individual nuclear spin resonances.

\begin{figure*}
  \includegraphics{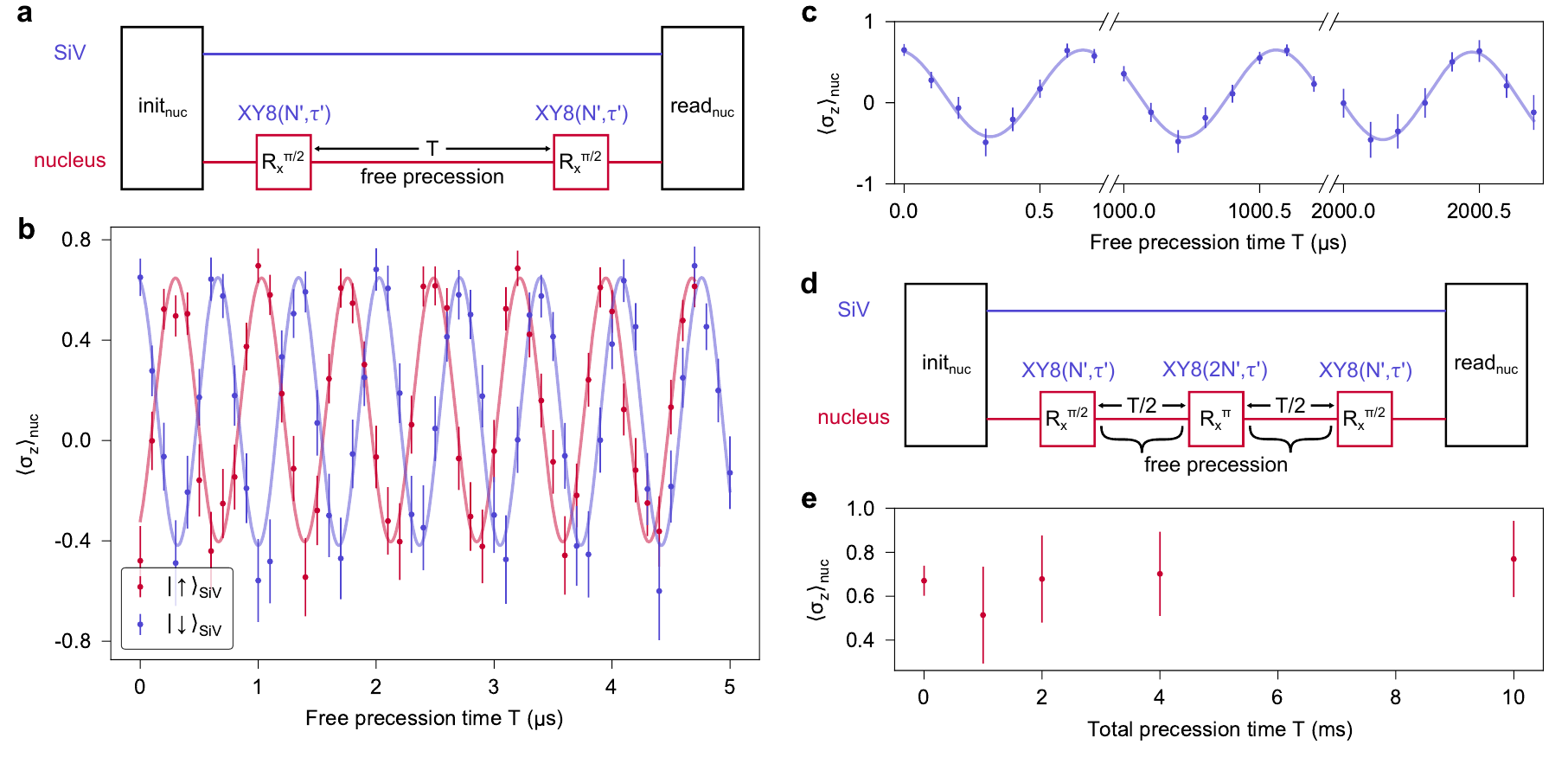}
  \caption{
    \label{fig4}
    \textbf{Mechanically driven Ramsey interferometry and spin echo of the \cnuc nuclear spin.}
    \textbf{a} Sequence for Ramsey interferometry.
    Two $\phi' = \pi/2$ rotations of the nuclear spin, corresponding to XY8 sequences with $N'=8$ and $\tau'=1.578\text{ $\mu$s}$, are separated by a free precession time $T$.
    \textbf{b} Mechanically driven Ramsey interferometry of the nuclear spin.
    The electron spin is initialized in $\ket\uparrow$ ($\ket\downarrow$) after the first $\pi/2$ nuclear rotation, corresponding to the red (blue) dots.
    The nuclear spin precesses at different rates depending on the state of the electron spin.
    The solid curves are sinusoidal fits, whose frequencies are consistent with the hyperfine coupling parameters determined earlier.
    \textbf{c} Ramsey oscillations measured at longer precession times.
    There is no decrease in the amplitude of the oscillations, indicating $T_2^* \gg 2\text{ ms}$ for the \cnuc nuclear spin.
    \textbf{d} Sequence for nuclear spin echo.
    The $\pi$-rotation of the nuclear spin is implemented using an XY8 sequence with $2N'=16$ and $\tau'=1.578\text{ $\mu$s}$.
    \textbf{e} The spin echo measurement shows no decay up to $10\text{ ms}$, indicating $T_2^\text{echo} \gg 10\text{ ms}$ for the \cnuc nuclear spin.
  }
\end{figure*}

If the inter-pulse delay is $\tau'=1.578\text{ $\mu$s}$, the XY8 sequence rotates the nuclear spin around the $X$-axis of the Bloch sphere by an angle dependent on the number of pulses $N'$ in the sequence.
By inserting such an XY8 sequence between the nuclear spin initialization and readout sequences (Fig.\ \ref{fig3}d), and varying $N'$, we perform coherent rotations of the nuclear spin, as demonstrated by the observation of Rabi oscillations shown in Fig.\ \ref{fig3}e.
We note that $N'=8$ corresponds to a $\pi/2$-rotation of the nuclear spin.

We then evaluate the coherence of the nuclear spin under mechanical control, using Ramsey interferometry.
We perform two $\pi/2$-rotations of the nuclear spin separated by a variable delay, whereby the nuclear spin is brought into a superposition state that precesses freely before being read out (Fig.\ \ref{fig4}a).
Immediately after the first $\pi/2$-rotation of the nuclear spin, the SiV is initialized in either $\ket\downarrow$ (red dots) or $\ket\uparrow$ (blue dots), which causes the nuclear spin to precess at a rate that depends on the state of the SiV spin, as shown in Fig.\ \ref{fig4}b, with precession frequencies $f_{\uparrow,\downarrow}^\text{Ramsey}=\sqrt{(f_\text{L} \pm A_\parallel/2)^2 + (A_\perp/2)^2}$.
This difference in precession frequencies confirms that we address a single nuclear spin.
We repeat the measurement for longer free precession times for a single orientation of the SiV spin, as displayed in Fig.\ \ref{fig4}c.
The amplitude of the Ramsey oscillations does not decrease up to precession times of $2\text{ ms}$, indicating that the nuclear spin retains its coherence above this timescale, $T_{2}^{*}\gg 2\text{ ms}$.
Next, we perform a spin echo sequence by inserting a nuclear $\pi$-rotation between the two $\pi/2$-rotations, as illustrated in Fig.\ \ref{fig4}d.
The $\pi$-rotation cancels the impact of any dephasing mechanism that is slower than the free precession timescale.
The nuclear spin state at the end of the sequence is plotted as a function of the total precession time in Fig.\ \ref{fig4}e and shows no sign of decay, indicating $T_2^\text{echo} \gg 10\text{ ms}$.
Due to the limited photon collection efficiency of our system, measurements for longer precession times are impractical, considering the experimental times required.
Nevertheless, our measurements confirm that the \cnuc nuclear spin retains long coherence times \cite{Bradley2019PRX, Nguyen2019PRB} under mechanical control.

In conclusion, we coherently couple surface acoustic waves to a single \cnuc nuclear spin using an SiV centre as an interface, and show that the nuclear spin retains its excellent coherence properties.
Future developments include integration with photonic structures such as tapered waveguides, which would increase the photon collection efficiency and enable single-shot spin readout \cite{Sukachev2017PRL}.
Furthermore, this method can be extended to other nuclear spins.
Using an SiV with a resident \sinuc nuclear spin enables the resolution of hyperfine energy levels \cite{Pingault2017NatComm}, which can be used to achieve entanglement between the SiV and the nuclear spin with a single $\pi$-pulse.
At the same time, strong coupling between the SiV spin and a single phonon could be achieved using a high quality factor, low mode volume mechanical resonator \cite{Burek2016Optica, Meesala2018PRB, Shao2019PRAppl}, thus enabling single-phonon spin gates.
This would enable the use of single nuclear spins as quantum memories for phononic states, with applications in metrology, quantum sensing \cite{Degen2017RMP}, quantum information processing \cite{Palomaki2013Nature, Schuetz2015PRX, Chu2017Science, Arrangoiz-Arriola2018PRX, Satzinger2018Nature}, and fundamental tests of quantum theory \cite{Berta2010NatPhys, Bassi2013RMP, Aspelmeyer2014RMP}.
Finally, diamond possesses a large Young's modulus allowing mechanical resonators to reach frequencies in the GHz range, thus making their mechanical ground states reachable with dilution refrigerators.
This, combined with recent progress in diamond nanofabrication, makes diamond an ideal platform for non-classical mechanics experiments, while taking advantage of the remarkable coherence properties of nuclear spins.

\section{Methods}

\subsection{Device fabrication}
We use $[100]$-cut, electronic grade single-crystal diamond samples synthesized by chemical vapour deposition (CVD) from Element Six Corporation.
Silicon ions ($^{28}\text{Si}^+$) are implanted on the top surface of the diamond at an energy of $150\text{ keV}$ and a density of $10^{10}\text{ cm}^{-2}$, introducing Si atoms over the entire surface at a depth of $100\pm 18\text{ nm}$ as determined by a SRIM simulation \cite{Ziegler2010NucInstMet}.
SiV centres are generated by a high-temperature ($1200^\circ\text{ C}$), high-vacuum annealing procedure followed by a tri-acid clean (1:1:1 sulphuric, perchloric and nitric acids).
A $1.4\text{ $\mu$m}$ aluminium nitride (AlN) layer is deposited on top of the diamond by RF sputtering.
IDTs are fabricated using electron beam lithography followed by evaporation of $75\text{ nm}$ of aluminium (Al).
The SAWs propagate along the $[110]$ direction, parallel to one of the edges of the sample.
Further details on the design and characterization of IDTs may be found in \cite{Maity2020NatComm}.

\subsection{Experimental setup}
Low temperature measurements are performed in a $^3\text{He}/^4\text{He}$ dilution refrigerator (Bluefors LD250) with a base temperature of $10\text{ mK}$.
The diamond sample is mounted on a plate which is thermally anchored to the mixing plate of the refrigerator.
The diamond is affixed with indium foil to a custom-made holder made of copper, which is located on top of an XY nanopositioner stack.
IDTs on the sample are wire-bonded to a printed circuit board (PCB) mounted on the holder, for electrical excitation of acoustic pulses.
A thermometer located close to the sample indicates a typical temperature of $50 - 200\text{ mK}$ during our experiments.
The sample is surrounded by a 3-axis superconducting vector magnet system used to apply a static magnetic field.
Above the sample, a cryogenic objective (attocube LT-APO/VISIR/0.82) with a numerical aperture of $0.82$ and a working distance of $0.65\text{ mm}$ is mounted on a Z nanopositioner.
A home-built optical microscope is used for optical excitation and fluorescence collection.
A tunable laser stabilized to an accuracy of $0.01\text{ pm}$ using feedback from a wavelength meter is used to resonantly excite optical transitions of individual SiVs.
Photons emitted by individual SiVs in the phonon side band (PSB) $> 740\text{ nm}$ are selected by an optical long-pass filter and sent to an avalanche photodiode (APD).

\subsection{Generation of optical and acoustic pulses}
Laser pulses at $737\text{ nm}$ are generated using an acousto-optic modulator (AOM).
The AOM is driven by a custom-built RF driver and is used in a double-pass configuration to achieve an extinction ratio $> 70\text{ dB}$ that is required to prevent unwanted optical spin pumping during long experimental sequences.
A time-correlated single photon counter (TCSPC) is used to record arrival times of photons emitted from the SiV.
An arbitrary waveform generator (AWG) controls the entire experimental sequence.
The 8-bit analog output of the AWG is used to drive the IDTs, while the two 1-bit marker outputs are used to control the AOM and trigger the TCSPC.
There is a small amount of drift between the clocks of the AWG and TCSPC, with a factor of about $1.25\times 10^{-5}$, which we correct for in our data processing.

\subsection{SiV spin initialization}
The SiV can be initialized in $\ket\uparrow$ or $\ket\downarrow$ by resonantly pumping the optical transitions $\downarrow\downarrow'$ or $\uparrow\uparrow'$ respectively.
For our experiments, after the acoustic $\pi$-pulse is calibrated, we initialize the SiV in $\ket\uparrow$ by optically pumping $\downarrow\downarrow'$, and in $\ket\downarrow$ by optically pumping $\downarrow\downarrow'$ followed by an acoustic $\pi$-pulse.
This allows keeping the wavelength of the laser resonant with the $\downarrow\downarrow'$ transition.
We insert an acoustic $\pi$-pulse between the final optical readout pulse of one experimental sequence and the optical initialization pulse of the next sequence, so that the SiV is in $\ket\downarrow$ before the initialization pulse.
This allows the photon counts during the initialization pulse to be used as a measure of the population in $\ket\downarrow$.

\subsection{SiV spin population measurement}
The SiV fluorescence is integrated for the first $100\text{ ns}$ of optical pumping.
The spin population of the SiV is measured as the ratio of the integrated fluorescence during the optical readout pulse to that during the optical initialization pulse.
The spin projection of the SiV is calculated as $\braket{\sigma_z} = 1 - 2P_\downarrow$, in which $P_\downarrow$ is the population in the $\ket\downarrow$ state.

\subsection{Nuclear spin rotations}
For the value of $\tau=1.569\text{ $\mu$s}$ used for nuclear initialization and readout, the rotation angle is calculated to be $\phi = 0.62 \pi$, with the conditional rotation axes being $\textbf{n}_\uparrow = (0.77,0, 0.64)$ and $\textbf{n}_\downarrow = (-0.69, 0, 0.72)$.

The $X$-rotations of the nuclear spin during the Rabi and Ramsey interferometry sequences are actually conditional rotations of the form $\mathcal R^\phi_{-\mathbf x,\mathbf x}$ for the inter-pulse time $\tau' = 1.578\text{ $\mu$s}$.
However, since the electron spin is initialized in $\ket\uparrow$ at the end of the initialization sequence, the nuclear spin only rotates about the $+X$-axis.

\begin{acknowledgments}
  The authors thank Neil Sinclair, Bartholomeus Machielse, Can Knaut, Mihir Bhaskar, David Levonian, Christian Nguyen, Hans Bartling, and Tim Taminiau for helpful discussions.
  This work was supported by NSF CIQM (Grant no.\ DMR-1231319), NSF ERC (Grant no.\ EEC-1941583), NSF RAISE TAQS (Grant no.\ ECCS-1838976), ARO MURI (Grant no.\ W911NF1810432), ONR (Grant no.\ N00014-20-1-2425), DOE (Grant no.\ DE-SC0020376), AFOSR (Grant no.\ FA9550-20-1-0105), and ONR MURI (Grant no.\ N00014-15-1-2761).
  B.P.\ acknowledges funding from the European Union's Horizon 2020 research and innovation programme under the Marie Sklodowska-Curie grant agreement no.\ 840968.
  This work was performed in part at the Harvard University Center for Nanoscale Systems (CNS); a member of the National Nanotechnology Coordinated Infrastructure Network (NNCI), which is supported by the National Science Foundation under NSF award no.\ ECCS-2025158.
  The computations in this paper were run on the FASRC Cannon cluster supported by the FAS Division of Science Research Computing Group at Harvard University.
\end{acknowledgments}

\section{Author contributions}
S.M.\ fabricated the devices with help from E.C.
S.M.\ and B.P.\ performed experimental measurements with help from G.J., M.C., L.S., and D.A.
S.M.\ and B.P.\ analyzed experimental data, and prepared the manuscript with help from all authors.
M.L.\ supervised the project.

\end{document}